\documentstyle[preprint,eqsecnum,aps]{revtex}
\begin{document}
\draft
%\preprint{HEP/123-qed}
\title{An Estimate of Charge Symmetry Breaking in Nuclear\\
Deep Inelastic Scattering}
\author{R.M. Davidson}
\address{NIKHEF,
Postbus 41882,
1009 DB Amsterdam,
The Netherlands\\
and\\
Department of Physics, Applied Physics and Astronomy\\
Rensselaer Polytechnic Institute,
Troy, NY 12180-3590\thanks{Present address}}
\author{M. Burkardt}
\address{Department of Physics\\
New Mexico State University\\
Las Cruces, NM 88003-0001}

%\date{\today}
\maketitle
\begin{abstract}
We estimate the magnitude of charge symmetry breaking effects in deep inelastic
scattering from nuclei. The resulting contribution to systematic
uncertainties in hadronic
determinations of $\sin^2\theta_W$ are found to be less than the current 
experimental accuracy, but
may be important in the analyses of
more precise future experiments. We expect the largest nuclear
charge symmetry breaking effects in the Paschos-Wolfenstein ratio
$R^-$, where the resulting uncertainty in the determination
of $\sin^2 \theta_W$ reaches $10^{-3}$.
\end{abstract}
%{\tt$\backslash$\string pacs\{\}} should always be input,
%even if empty.}
\narrowtext
\section{Introduction}
With the systematic improvement in hadronic determinations
of $x_w \equiv \sin^2 \theta_W$ from deep inelastic scattering (DIS)
experiments \cite{arroyo}, it becomes increasingly important to get also the 
theoretical systematic
uncertainties under control. While many theoretical and experimental
uncertainties cancel when one considers the Paschos-Wolfenstein ratios 
\cite{wolf},
\begin{equation}
R^{\pm} = \frac{ d\sigma^\nu_{NC}/dy \pm d\sigma^{\overline{\nu}}_{NC}/dy}
{ d\sigma^\nu_{CC}/dy \pm d\sigma^{\bar{\nu}}_{CC}/dy}
\label{eq:rpm}
\end{equation}
and
\begin{eqnarray}
R^\nu &=&\frac{d\sigma^\nu_{NC}/dy}{d\sigma^\nu_{CC}/dy} ,
\nonumber\\
R^{\bar{\nu}} &=&\frac{d\sigma^{\bar{\nu}}_{NC}/dy}{d\sigma^{\bar{\nu}}_{CC}/dy} 
,
\label{eq:rnu}
\end{eqnarray}
there are still several important corrections to these ratios that
arise, for example, from the strange and charm sea, radiative
corrections \cite{amaldi} and charge symmetry breaking (CSB)
effects in the nucleon \cite{sather}.
While CSB effects for parton
distributions in the nucleon have been considered \cite{sather,goldman},
possible CSB effects arising from the Coulomb contribution to the nuclear 
binding
energy have so far been neglected.
In this Letter, we make a simple estimate of the size of such nuclear
charge symmetry breaking effects on the Paschos-Wolfenstein ratios and
concomitantly their effect on the determination of $x_w$ from these ratios.

\section{The model}
Let us first recall some formulae relevant for the determination of $x_w$
from neutrino-nucleus deep inelastic scattering.
Given the momentum fractions $\langle q \rangle$ carried by the various flavors 
of
quarks in the target, one finds for the inclusive cross sections \cite{proy}
\begin{eqnarray}
{d\sigma^\nu_{CC} \over dy }=
\frac{G_F^2 s}{\pi} \frac{M_W^4}{\left(M_W^2+Q^2\right)^2} \left\{ 
\left(1-y+\frac{y^2}{2}\right)
\langle d+s+\bar{u}+\bar{c} \rangle
+y\left(1-\frac{y}{2}\right) \langle d+s-\bar{u} -\bar{c} \rangle \right\}
\nonumber \\
{d\sigma^{\bar{\nu}}_{CC} \over dy }=
\frac{G_F^2 s}{\pi} \frac{M_W^4}{\left(M_W^2+Q^2\right)^2} \left\{ 
\left(1-y+\frac{y^2}{2}\right)
\langle u+c+\bar{d}+\bar{s} \rangle
-y\left(1-\frac{y}{2}\right) \langle u+c-\bar{d} -\bar{s} \rangle \right\}
\nonumber \\
{d\sigma^{\nu ,\bar{\nu}}_{NC} \over dy }=
\frac{G_F^2 s}{\pi} \frac{M_Z^4}{\left(M_Z^2+Q^2\right)^2} \left\{ 
\left(1-y+\frac{y^2}{2}\right) \left[ \alpha_{1+}
\langle u+c+\bar{u}+\bar{c} \rangle
+ \alpha_{1-} \langle d+s+\bar{d}+\bar{s} \rangle \right] \right. \nonumber \\
\pm \left. y\left(1-\frac{y}{2}\right) \left[ \alpha_{2+}
\langle u+c-\bar{u} -\bar{c} \rangle + \alpha_{2-}
\langle d+s-\bar{d} -\bar{s} \rangle \right]
\right\} \; ,
\end{eqnarray}
where $y\equiv 
\left[\left(E_{initial}-E_{final}\right)/E_{initial}\right]_{lab}$ and the
subscripts `CC' and `NC' refer to the charged and neutral currents, 
respectively.
In the last of these equations, the `+' refers to $\nu$ and the `-' to 
$\bar{\nu}$, and
\begin{eqnarray}
\alpha_{1+} &=& { 1 \over 4} -{ 2 \over 3} x_w + {8 \over 9 } x_w^2 
\;\;\;\quad\quad\quad
\alpha_{2+} = { 1 \over 4} - {2 \over 3} x_w \nonumber \\
\alpha_{1-} &=& {1 \over 4 } -{1 \over 3} x_w + {2 \over 9} x_w^2 
\;\;\;\quad\quad\quad
\alpha_{2-} = { 1 \over 4} - {1 \over 3} x_w \; .
\end{eqnarray}

Assuming $\langle s \rangle$ = $\langle \bar{s} \rangle$ and
$\langle c \rangle$ = $\langle \bar{c} \rangle$, the Paschos-Wolfenstein ratios
may be written as\footnote{We consider kinematics where Q$^2$ $<<$ $M_W^2$.}
\begin{eqnarray}\label{eq:y}
R^- &=& { f_2 (y) \left[ (\alpha_{2+} + \alpha_{2-}) 
\langle Q_{+} \rangle -
(\alpha_{2+}-\alpha_{2-}) \langle Q_{-} \rangle \right] 
\over f_2 (y) \langle Q_{+} \rangle + f_1 (y) \langle Q_{-} \rangle } \nonumber 
\\
R^+ &=& { f_1 (y) \left[ (\alpha_{1+}+\alpha_{1-}) \langle Q_{+}+2\bar{Q}_1
\rangle -(\alpha_{1+}-\alpha_{1-}) \langle Q_{-}+\bar{Q}_{2} \rangle \right]
\over f_1 (y) \langle Q_{+} +2\bar{Q}_{1} \rangle + f_2 (y) \langle
Q_{-} +2 \bar{Q}_{2} \rangle } \nonumber \\
R^{\nu , \bar{\nu}} &=& \left\{ f_1 (y) \left[ \alpha_{1+} 
\langle Q_{+} -Q_{-}
+4\bar{Q}_{uc} \rangle +\alpha_{1-} \langle Q_{+} + Q_{-}
+4\bar{Q}_{ds} \rangle \right] \right. \nonumber \\
 &\pm& \left. f_2 (y) \left[ \alpha_{2+} \langle Q_{+}
-Q_{-} \rangle +\alpha_{2-} \langle Q_{+} +Q_{-} \rangle \right] \right\}
\nonumber \\
 &\times& \left\{ f_1 (y) \langle Q_{+} 
\pm Q_{-} + 2\bar{Q}_1 \rangle \pm \langle Q_{+}
\pm Q_{-} \pm 2\bar{Q}_2 \rangle \right\} ^{-1} \; ,
\end{eqnarray}
where $ \langle Q_{\pm} \rangle$ = $\langle d_v \pm u_v \rangle$, 
$\langle \bar{Q}_{uc}\rangle$ = $\langle \bar{u}+\bar{c} \rangle$,
$\langle \bar{Q}_{ds}\rangle$ = $\langle\bar{d}+\bar{s}\rangle$, 
$\langle\bar{Q}_1 \rangle$ = $\langle\bar{d}+\bar{u}+\bar{s}+\bar{c}\rangle$
and $\langle\bar{Q}_2 \rangle$ = 
$\langle\bar{d}+\bar{s}-\bar{u}-\bar{c}\rangle$.
Also, $f_1 (y)$ =
$1-y+y^2 /2$ and $f_2 (y)$ = $y-y^2 /2$. In the last of the equations, `+' 
refers
to $\nu$ and `-' to $\bar{\nu}$.

As the above ratios, Eq.~(\ref{eq:y}),
are dominated by the valence quark contributions, $\langle Q_{+} \rangle$,
the scale of CSB effects in an isoscalar target is set by the ratio
$\langle Q_{-} \rangle / \langle Q_{+} \rangle$. Evidently, for an
isoscalar target $\langle Q_{-} \rangle$ vanishes in the absence of CSB effects.
In order to obtain a simple and transparent estimate for this ratio, and hence
nuclear CSB effects in DIS, we employ the convolution approach of
Dunne and Thomas \cite{dunne}, where the momentum fractions of a
quark in the nucleus, $\langle q \rangle_A$,
are\footnote{As we are ultimately interested in ratios, the ``rescaling" effects
may be neglected.}
\begin{equation}
A \langle q \rangle_A = \sum_i \left\{
n_{ip}\frac{M_{ip}}{{M_p}} \langle q \rangle_p
+ n_{in}\frac{M_{in}}{{M_n}} \langle q \rangle_n \right\},
\label{eq:dunne}
\end{equation}
where $A$ is the nuclear mass number, $M_{p(n)}$ is the proton (neutron) mass,
$n_{ip(n)}$ is the proton (neutron) occupation number for the i-th
shell, $M_{ip(n)}= M_{p(n)}-\varepsilon_{ ip(n)}$ and
$\varepsilon_{ ip(n)}$ is the proton (neutron) separation energy
for the i-th shell.
There are some difficulties with the overall momentum sum rule in the
convolution approach, so one should not consider Eq.~(\ref{eq:dunne})
an exact formula. However, for the purpose of this work, namely obtaining
a rough estimate of the size of nuclear charge symmetry breaking in the
Paschos-Wolfenstein ratios, such subtleties can be ignored.

In this approach, we thus obtain
\begin{equation}
\langle Q_\pm \rangle = \sum_i
n_{ip} \frac{M_{ip}}{M_p} \langle u_v \pm d_v \rangle_p
+ n_{in} \frac{
M_{in}}{M_n} \langle u_v \pm d_v \rangle_n .
\end{equation}
As we are interested in {\it nuclear} CSB effects, we
assume charge symmetry for the nucleon and obtain
\begin{equation}\label{qmpr}
\frac{\langle Q_-\rangle }{ \langle Q_+\rangle}
=
\frac{ \langle u_v - d_v \rangle_p }
{ \langle u_v + d_v \rangle_p }
\frac{ \left\{ \sum_i n_{in} M_{in}
- n_{ip} M_{ip} \right\} }
{ \left\{ \sum_i n_{in} M_{in}
+ n_{ip} M_{ip} \right\} } \; .
\end{equation}

\section{Numerical Results}
Since the $y$-dependence in Eqs.~(\ref{eq:y}) is not very large except for 
$R^{-}$
at very small $y$, we consider from now on the ratios of
the $y$-integrated cross sections
in order to get some overall quantitative understanding
about the relevance of CSB in these ratios.
Since the CSB contribution to the Paschos-Wolfenstein ratios,
Eqs.~(\ref{eq:rpm}), (\ref{eq:rnu}),
are rather small, it is not necessary to include other corrections
(charm, etc.) when calculating the CSB corrections. To estimate $M_{ip}$ and
$M_{in}$, we take the separation energies given in Ref.~\cite{dunne} and 
estimate
the Coulomb correction, V$_c$, by considering the difference in masses between
the nucleus of interest the nucleus with one proton or neutron removed. For 
example,
to estimate V$_c$ for $^4$He, we compare its mass with the masses of $^3$He and
$^3$H. For heavier nuclei such as $^{40}$Ca, this comparison actually gives the
difference in separation energies for the outer most shell. However, detailed
calculations of the separation energies including the Coulomb interaction show
that the difference in separation energies is largely independent of the shell.
Given V$_c$, we then take $M_{ip}$ = $M_{iN}$ + V$_c$/2 and $M_{in}$ =
$M_{iN}$ - V$_c$/2, where the $M_{iN}$ are given in Ref.~\cite{dunne}.
Our results for V$_c$ and $\langle Q_{-} \rangle / \langle Q_{+} \rangle $
are given in Table I, where we have taken
$\langle d_v \rangle / \langle u_v \rangle$ = 0.4434 \cite{owens}.

The scale of the results in Table I %\ref{table1} 
can easily be understood by noting the the sum in the
denominator of Eq.~(\ref{qmpr}) is nearly equal to $AM$ and the sum in the
numerator is ZV$_c$. Taking, for simplicity, the SU(6) result of
$\langle d_v \rangle / \langle u_v \rangle$ = 1/2, one obtains
\begin{equation}
{ \langle Q_{-} \rangle \over \langle Q_{+} \rangle } \approx
{V_c \over 6M} \; .
\end{equation}
As an additional check on these results, the sum in the numerator in 
Eq.~(\ref{qmpr})
is a sum over the Coulomb energies of the individual protons, and is thus equal
to twice the Coulomb energy of the nucleus, E$_c$. Assuming a uniform
charge density and a nuclear radius that scales like $R$ = 1.2fm$A^{1/2}$, we
find for isoscalar nuclei
\begin{equation}
{ \langle Q_{-} \rangle \over \langle Q_{+} \rangle } \approx
- {\langle u_v -d_v \rangle _p \over \langle u_v +d_v \rangle _p }
{ 0.57 {\rm MeV} \over M} Z^{2/3} \; .
\end{equation}
The numbers obtained with this latter formula are in good agreement with those
in Table I, with the exception of $^4$He.

Given the numbers in Table I, it is now easy to determine the nuclear CSB 
effects
on the ratios. In particular, we are interested in how much the ratios change
when the CSB effects are included. Thus, we show in Table II the results for
$\delta R$ = $R_c$ - $R_0$ where $R_c$ is the ratio including CSB effects. We 
also
show the results for $\delta_w$ = $x_{w'}$ -$x_w$ \cite{amaldi},
where $x_{w'}$ is determined
by demanding that $R$ does not change when CSB effects are included, i.e.
$R_0(x_w)$ = $R_c(x_{w'})$.
For calculations, we use $x_w$ = 0.23, and
thus have $R_0^{-}$ = 0.2700, $R_0^{+}$ = 0.3288, $R_0^{\nu}$ = 0.3092, and
$R_0^{\bar{\nu}}$ = 0.3876.

Evidently, of the Paschos-Wolfenstein ratios it is $R^{-}$ that is most 
influenced
by CSB effects. For the other ratios, the  CSB effect is roughly 1/2 the size
of the $\langle c \rangle$ contribution\footnote{Based on Owens'
\protect\cite{owens} parameterization at Q$^2$ = 10 GeV$^2$.} for $^{28}$Si
and $^{40}$Ca and about 1/6 the size of the $\langle c \rangle$ contribution
for $^4$He and $^{12}$C. For $\delta_w$, it is $\delta_w^{\bar{\nu}}$ that
is most affected by CSB affects, but this contribution is at most 1/3 of the
$\langle c \rangle$ contribution. Compared to
the $\langle c \rangle$ contribution, the CSB effect in $\delta_w^{\nu}$ is 
quite
important, being about 2/3 the size of it for $^{40}$Ca.

While $R^{-}$ is certainly the best ratio to look for CSB effects, particularly
at small $y$, we remind the reader that we have assumed
$\langle s \rangle$ = $\langle \bar{s} \rangle$. If this assumption is relaxed, 
there
will be contributions to $R^{-}$ of the form $\langle s - \bar{s} \rangle$.
\footnote{For a discussion of the $s-\bar{s}$ asymmetry in nucleons, 
see e.g. Ref. \cite{ssbar} and references therein.} Although
this is certainly small, its size needs to be known before one can
get information about $\langle Q_{-} \rangle$ from $R^{-}$.
$R^\nu$ receives a similarly large CSB correction, but $R^\nu$ is 
more strongly 
affected by the strange and charm corrections, relative to which nuclear
CSB plays only a minor role.

\section{summary}
We have investigated the contribution of nuclear binding effects
to charge symmetry breaking for parton momentum fractions
in isoscalar targets.
Due to the Coulomb repulsion, protons are less strongly bound in
nuclei and thus carry a larger fraction of the nucleus' momentum.
Up to a factor of 3 suppression, which results from a partial cancelation
between quarks of the same flavor in protons and neutrons, this
translates directly into $u$ quarks carrying more momentum than $d$ quarks
due to the Coulomb energy in nuclei.

Naively, one would expect nuclear charge symmetry breaking effects
to be of the order of the Coulomb energy of the nucleus
divided by its mass, i.e.~an effect of the order of 1\% for
larger nuclei. Actually, the average effect per nucleon 
is only 0.5\%, since only protons, but not
neutrons, are affected.
However, the 0.5\% charge symmetry breaking in the distribution
of nucleons in nuclei gets ``diluted'' roughly by abovementioned factor of
three when one considers quarks. As a result, the net charge symmetry
breaking at the quark level is only about 0.1-0.2\%, i.e.~almost
an order of magnitude smaller than the most naive estimate.

Of course, our model is very naive and one should thus not view
our result as a prediction, but rather as an estimate for the
systematic uncertainties that arise due to nuclear
charge symmetry breaking effects.
Depending on the target and on
which of the Paschos-Wolfenstein ratios one uses for the
determination of $\sin^2\theta_W$, the resulting systematic
error due to charge symmetry breaking nuclear effects is thus
of the order of $0.1-0.5\%$, i.e.~slightly smaller than the expected
effects due to charm quarks in the target. As a result, nuclear isospin
symmetry breaking effects are still too small to be detected,
but this may soon change as more precise experimental data become
available.

We predict the largest nuclear CSB corrections in the ratio $R^-$.
Incidentally, most other corrections to the naive Paschos-Wolfenstein
result cancel in this particular ratio. Thus for $R^-$, nuclear CSB might
be, besides CSB in the nucleon and a possible charge conjugation
asymmetry in the strange sea \cite{ssbar},
the largest nonperturbative correction to the naive Paschos-Wolfenstein 
ratio.

\acknowledgements
RMD was supported in parts by the U.S.~Dept.~of Energy
grant number DE-FG02-88ER40448, to N.C.~Mukhopadhyay,
and the Foundation for Fundamental Research on Matter (FOM) and the
National Organization for Scientific Research (NWO) (The Netherlands). MB was 
supported by the U.S. Dept.~of Energy under contract DE-FG03-96ER40965
and in part by TJNAF.

\begin{table}
\label{table1}
\caption{Our estimate of the average difference between proton and neutron 
separation
energies due to the Coulomb interaction, V$_c$, and
the ratio $\langle Q_{-} \rangle / \langle Q_{+} \rangle $ for various
N=Z nuclei.}
\begin{center}
\begin{tabular}{|r|c|c|} \hline
Nucl. & V$_c$ (MeV) & $\langle Q_{-} \rangle / \langle Q_{+} \rangle $  \\ 
\hline
 $^4$He & 0.77 & -0.0007661 \\
 $^{12}$C & 2.77 & -0.0008523 \\
 $^{28}$Si & 5.60 & -0.0014512 \\
 $^{40}$Ca & 7.31 & -0.0018153 \\ \hline
\end{tabular}
\end{center}
\end{table}

\begin{table}
\label{table2}
\caption{The results for $\delta R$ and $\delta_w$ (see text) due to
nuclear CSB effects for various N=Z nuclei.}
\begin{center}
\begin{tabular}{|r|c|c|c|c|c|c|c|c|} \hline
Nucl. & $\delta$R$^{-}$ & $\delta _w^{-}$
& $\delta$R$^{+}$ & $\delta _w^{+}$
& $\delta$R$^{\nu}$ & $\delta _w^{\nu}$
& $\delta$R$^{\bar{\nu}}$ & $\delta _w^{\bar{\nu}}$
\\ \hline
 $^4$He & 0.0002 & 0.0002 & 0.0001 & 0.0001 & 
 0.0000 & 0.0001 & -0.0001 & -0.0006 \\
 $^{12}$C & 0.0004 & 0.0004 & 0.0001 & 0.0002 & 
 0.0001 & 0.0003 & -0.0001 & -0.0011 \\
 $^{28}$Si & 0.0007 & 0.0007 & 0.0002 & 0.0003 & 
 0.0003 & 0.0005 & -0.0003 & -0.0018 \\
 $^{40}$Ca & 0.0008 & 0.0008 & 0.0002 & 0.0003 & 
 0.0003 & 0.0006 & -0.0003 & -0.0023 \\ \hline
\end{tabular}
\end{center}
\end{table}


\begin{references}
\bibitem{arroyo} K.S. McFarland et al., Talk presented at 31st Rencontres de 
Moriond,
Electroweak Interactions \& Unified Theories, Les Arcs, France, 16-23 Mar 1996.
e-Print Archive: hep-ex/9608007
\bibitem{wolf} E.A. Paschos and L. Wolfenstein, Phys.\ Rev.\ D\ {\bf 7}, 91 
(1973).
\bibitem{proy} P. Roy, Lectures given at the 11th Mt. Sorak Symposium (July 
1992),
Sok-cho, Korea.
\bibitem{amaldi} U. Amaldi et al., Phys.\ Rev.\ D\ {\bf 36}, 1385 (1987).
\bibitem{sather} E. Sather, Phys.\ Lett.\ B\ {\bf 274}, 433 (1992).
\bibitem{goldman} C.J. Benesh and T. Goldman, e-Print Archive: nucl-th/9609024.
\bibitem{measure} P.G. Reutens et al., Phys.\ Lett.\ B\ {\bf 152}, 404 (1985).
\bibitem{dunne} G.V. Dunne and A.W. Thomas, Nucl.\ Phys.\ A\ {\bf 455}, 701 
(1986).
\bibitem{owens} J.F. Owens, Phys.\ Lett.\ B\ {\bf 226}, 126 (1991).
\bibitem{ssbar} A.I. Signal and A.W. Thomas, Phys.\ Lett.\ B\ {\bf 191}, 205 
(1987); M. Burkardt and B. Warr, Phys.\ Rev.\ D\ {\bf 45}, 958 (1992); S.J. 
Brodsky and B.-Q. Ma, Phys.\ Lett.\ B\ {\bf 381}, 317 (1996).
\end{references}
\end{document}